\documentclass[12pt,a4paper]{article}
\usepackage{a4wide}
\usepackage{latexsym}
\usepackage{epsf}
\begin{document}

\begin{flushright}
\small
IFT-UAM/CSIC-99-1\\
{\bf hep-th/9901078}\\
January $18$th, 1999
\normalsize
\end{flushright}

\begin{center}

%title

\vspace{.7cm}

{\Large {\bf The D8-Brane Tied up: String and Brane Solutions 
in Massive Type~IIA Supergravity}}

\vspace{.7cm}

%authors

{\bf\large Bert Janssen}${}^{\spadesuit\clubsuit}$
\footnote{E-mail: {\tt Bert.Janssen@uam.es}},
{\bf\large Patrick Meessen}${}^{\spadesuit}$
\footnote{E-mail: {\tt Patrick.Meessen@uam.es}} 
{\bf\large and Tom\'as Ort\'{\i}n}${}^{\spadesuit\clubsuit}$
\footnote{E-mail: {\tt tomas@leonidas.imaff.csic.es}}
\vskip 0.4truecm

${}^{\spadesuit}$\ {\it Instituto de F\'{\i}sica Te\'orica, C-XVI,
Universidad Aut\'onoma de Madrid \\
E-28049-Madrid, Spain}

\vskip 0.2cm

${}^{\clubsuit}$\ {\it I.M.A.F.F., C.S.I.C., Calle de Serrano 113 bis\\ 
E-28006-Madrid, Spain}

\vspace{.7cm}

%%%%%%%%%%%%%%%%%%%%%%%%%%%%%%%%%%%%%%%%%%%%%%%%%%%%%%%%%%%%%%%%%%%%%%
{\bf Abstract}

\end{center}

\begin{quotation}

\small

We present two new solutions of Romans' massive type~IIA supergravity
characterized by the two non-trivial massive potentials of Romans'
theory: the NSNS 2-form and the RR 7-form. They can be interpreted
respectively as the intersection of a fundamental string and a
D8-brane over a D0-brane and the intersection of a D6-brane with a
D8-brane over a NSNS5-brane.  The D8-brane manifests itself through
the mass parameter and in the massless limit one recovers the standard
fundamental string and D6-brane solutions.

Although these solutions do not have the usual form for BPS bound states
at threshold and each of them involves 3 objects, both of them preserve
$1/4$ of the supersymmetries. 

\end{quotation}

\vspace{1cm}

\newpage

\pagestyle{plain}

%%%%%%%%%%%%%%%%%%%%%%%%%%%%%%%%%%%%%%%%%%%%%%%%%%%%%%%%%%%%%%%%%%%%%%

\section*{Introduction}

Supergravity theories (and specially
SUEGRAs\footnote{SUper-Extended-GRAvity theories.}) are the low-energy
effective actions of string theories and therefore harbor fields
corresponding to the string massless modes. In particular they contain
$p+1$-form potentials whose sources can be identified with perturbative
and non-perturbative string theory $p$- and
$\tilde{p}=d-p-4$-dimensional objects: the fundamental string, the
solitonic 5-brane and the Dirichlet D$p$-branes. 

Type~IIA superstring theory contains D$p$-branes with $p=0,2,4,6,8$
\cite{kn:P}. However, the standard type~IIA SUEGRA
\cite{kn:HW,kn:HN,kn:GP} only contains RR potentials for the first four
cases. It was later realized \cite{kn:PW,kn:BRGPT} that since an 8-brane
couples to a 9-form potential which can be dualized into a constant
parameter (``-1-form field strength''), the low-energy theory of the
type~IIA superstring is Romans' massive type~IIA theory \cite{kn:Ro2},
until then considered an exotic deformation of the standard theory. 

Romans' theory contains a constant parameter $m$ with dimensions of mass
which can be interpreted as the Hodge dual of the 10-form field strength
associated to a D8-brane \cite{kn:BRGPT}.  The D8-brane carries no
dynamical degrees of freedom and should be interpreted as a background
for the type~IIA theory. This is however very non-trivial background:
the parameter not only appears in the action through a kinetic-like term

\begin{displaymath}
\int d^{10}x\, \sqrt{|g|}\, \left[-{\textstyle\frac{1}{2}}m^{2}\right]\, ,
\end{displaymath}

\noindent with the (in)dependence of the dilaton characteristic
of RR potentials (in the string frame), but also appears as a mass
parameter for the NSNS 2-form.

Romans' massive type~IIA theory has many mysterious features. In
particular, its 11-dimensional origin (and that of the D8-brane) cannot
be the standard 11-dimensional supergravity and one has to appeal to
non-covariant generalizations associated to backgrounds with special
isometric directions \cite{kn:BLO,kn:BS,kn:MO} (see also
\cite{kn:Hu2,kn:Hu5}). 

The physical reason why the NSNS 2-form $B$ gets a mass in this
background is not well understood either although this is the most
notorious feature of this theory, perhaps because it has not been fully
appreciated.  Actually, the mass term for $B$ has been ignored in the
search for supersymmetric solutions describing the intersection of
fundamental strings or solitonic 5-branes with
D8-branes\footnote{M-brane and D$p$-brane intersections were first
discussed in Refs.~\cite{kn:PT,kn:BBJ,kn:GKT}.}. 

Furthermore, in Ref.~\cite{kn:BLO} a previously ignored mass term for
the RR 7-form potential $C^{(7)}$ was shown to occur. 

In this letter we will present two supersymmetric solutions of massive
type~IIA SUEGRA for which these two mass terms do not vanish.  The
first solution which we will call {\it massive string} has a non-trivial
$B$ field (plus metric and dilaton). It preserves $1/4$ of the
supersymmetries and contains an arbitrary harmonic function allowing for
the description of several objects in equilibrium.  In the massless
limit it becomes the fundamental string solution \cite{kn:DGHR} and it
will be interpreted as the intersection of a fundamental string with a
D8-brane, the former occupying the only direction orthogonal to the
latter. It can be argued that there is a D0-brane present at the
intersection.  The existence of this configuration was demonstrated in
Ref.~\cite{kn:Sa}. 

The second solution, which we will call {\it massive D6-brane} has a
nontrivial $C^{(7)}$ field (plus metric and dilaton. Consistency of the
equations of motion requires the presence of a non-trivial $B$.  This
solution also preserves $1/4$ of the supersymmetries and contains an
arbitrary harmonic function allowing for the description of several
objects in equilibrium.  In the massless limit the $B$ field vanishes
and the solution becomes the D6-brane solution and it will be
interpreted as the intersection of a D6-brane with a D8-brane over 5
spacelike directions.  However, the presence of the $B$ field in the
massive case will be interpreted as a solitonic 5-brane living in the
intersection.

%%%%%%%%%%%%%%%%%%%%%%%%%%%%%%%%%%%%%%%%%%%%%%%%%%%%%%%%%%%%%%%%%%%%%%

\section{Massive Type~IIA SUEGRA}

Here we give the bosonic equations of motion and the fermionic
supersymmetry transformation laws of massive type~IIA SUEGRA in
the string frame including the dual RR and NSNS potentials. Many of
the general expressions are also valid for the IIB theory. Due to the
explicit occurrence of potentials in the action, they can only be
dualized on-shell. The dual potentials are defined by the relations
between field strengths

\begin{equation}
\left\{
\begin{array}{rcl}
G^{(10-n)} & = & (-1)^{\left[n/2\right]} {}^{\star}G^{(n)}\, ,  \\
& & \\
H^{(7)} & = & e^{-2\phi} {}^{\star} H\, ,\\
\end{array}
\right.
\end{equation}

\noindent plus the Bianchi identities 

\begin{equation}
\left\{
\begin{array}{rcl}
dG -H\wedge G & = & 0\, ,\\
& & \\
d H & = & 0\, ,\\
& & \\
dH^{(7)} +\frac{1}{2} {}^{\star} G \wedge G & = & 0\, ,\\
\end{array}
\right.
\end{equation}

\noindent and the equations of motion 

\begin{equation}
\left\{
\begin{array}{rcl}
d{}^{\star}G +H\wedge G & = & 0\, ,\\
& & \\
d \left(e^{-2\phi} {}^{\star} H \right) 
+\frac{1}{2} {}^{\star} G \wedge G & = & 0\, ,\\
& & \\
d\left( e^{2\phi} {}^{\star} H^{(7)}  \right)& = & 0\, ,\\
\end{array}
\right.
\end{equation}

\noindent where we are using the notation \cite{kn:Douglas,kn:GHT,kn:BLO}
in which which forms of different degrees are formally combined into a
single entity:

\begin{equation}
\left\{
\begin{array}{rcl}
C & = & C^{(0)} + C^{(1)} + C^{(2)} +\ldots\, , \\
& & \\
G & = & G^{(0)} + G^{(1)} + G^{(2)} +\ldots\, .\\
\end{array}
\right.
\end{equation}

These expressions are valid both for the type~IIB and for the massive
type~IIA theory if one selects respectively odd and even rank and odd
rank RR differential form field strengths and one makes the
identification

\begin{equation}
G^{(0)}=m\, .  
\end{equation}

The field strengths that correspond to these Bianchi identities and
equations of motion are given by

\begin{equation}
\left\{
\begin{array}{rcl}
G & = & dC -H\wedge C +m e^{B}\, ,\\
& & \\
H^{(7)} & = & dB^{(6)} -m C^{(7)} 
-\frac{1}{2} \sum_{n=1}^{n=4} {}^{\star} G^{(2n+2)}\wedge C^{(2n-1)}\, ,\\
& & \\
{\cal H}^{(7)} & = &  d{\cal B}^{(6)} 
+\frac{1}{2} \sum_{n=1}^{n=4} {}^{\star} G^{(2n+3)}\wedge C^{(2n)}\, ,\\
\end{array}
\right.
\end{equation}

\noindent where calligraphic fields belong to the IIB theory.

These equations have to be supplemented by the  dilaton 
equation of motion

\begin{equation}
R +4\left(\partial\phi\right)^{2}  -4\nabla^{2}\phi 
+{\textstyle\frac{1}{2\cdot 3!}}H^{2}  = 0\, ,
\end{equation}

\noindent and the Einstein equation of motion (where we have already
eliminated $R$ with the use of the dilaton equation of motion)

\begin{equation}
R_{\mu\nu} -2\nabla_{\mu}\nabla_{\nu}\phi 
+{\textstyle\frac{1}{4}} H_{\mu}{}^{\rho\sigma} H_{\nu\rho\sigma}
-{\textstyle\frac{1}{4}}e^{2\phi}\sum_{n} 
{\textstyle\frac{(-1)^{n}}{(n-1)!}} T^{(n)}{}_{\mu\nu}\, ,
\end{equation}

\noindent where $T^{(n)}{}_{\mu\nu}$ are the energy-momentum tensor
of the RR fields:

\begin{equation}
T^{(n)}{}_{\mu\nu} = G^{(n)}{}_{\mu}{}^{\rho_{1}\cdots\rho_{n-1}}
G^{(n)}{}_{\nu\rho_{1}\cdots\rho_{n-1}}
-{\textstyle\frac{1}{2n}} g_{\mu\nu} G^{(n)\, 2}\, ,
\end{equation}

\noindent and, in particular

\begin{equation}
T^{(0)}{}_{\mu\nu} = -{\textstyle\frac{1}{2}}m^{2} g_{\mu\nu}\, .
\end{equation}

These equations are also valid both for the type~IIA and IIB theories.
Observe that the contributions of the energy-momentum tensors of dual
fields add up, except in the $n=5$ case.

Let us now focus on the (massive) type~IIA theory. The supersymmetry
transformation law for the gravitino and dilatino are\footnote{We work
  with real 32-component ad purely imaginary gamma matrices satisfying
  $\{\Gamma^{a},\Gamma^{b}\}=2\eta^{ab}$ where $\eta^{ab}$ has mostly
  minus signature. Finally,
  $\Gamma_{11}=-\hat{\Gamma}^{0}\ldots\hat{\Gamma}^{9}$.}

\begin{equation}
\label{eq:IIAsusyrules2}
\left\{
\begin{array}{rcl}
\delta_{\epsilon} \psi_{\mu} & = & 
\left\{  
\partial_{\mu} -\frac{1}{4} \left(\not\!\omega_{\mu} 
+\frac{1}{4}\Gamma_{11}\not\!\! H_{\mu} 
+\frac{1}{2\cdot 7!}e^{2\phi}\Gamma_{\mu\nu_{1}\cdots\nu_{7}} 
H^{(7)\, \nu_{1}\cdots\nu_{7}}\right)
\right\} \epsilon \\
& & \\ 
& &  
+\frac{i}{16} e^{\phi} \Sigma_{n=0}^{n=4} \frac{1}{(2n)!}
\not\! G^{(2n)} \Gamma_{\mu} 
\left( -\Gamma_{11} \right)^{n}\epsilon\, , \\
& & \\
\delta_{\epsilon}\lambda & = &   
\left[\not\!\partial\phi
+\frac{1}{4}\left(\frac{1}{3!}\Gamma_{11}\not\!\! H
-\frac{1}{7!}e^{2\phi}\!\!\not\!\!H^{(7)}\right)\right]
\epsilon 
+ \frac{i}{8} e^{\phi} \sum_{n=0}^{n=4} \frac{5-2n}{(2n)!} 
\not\! G^{(2n)} \left(-\Gamma_{11} \right)^{n} 
\epsilon\, .\\
\end{array}
\right.
\end{equation}

As explained in the introduction, the mass parameter occurs in the
form of a cosmological constant (or, in the Einstein frame, of an
unbound potential for the dilaton). Furthermore, the field strengths
of $C^{(1)}$ and $B^{(6)}$ contain the terms

\begin{equation}
\left\{
\begin{array}{rcl}
G^{(2)} & = & d C^{(1)} +mB\, ,\\
& & \\
H^{(7)} & = & d B^{(6)} -mC^{(7)} +\ldots\\ 
\end{array}
\right.
\end{equation}

\noindent associated to these terms there are two {\it massive gauge
transformations} 

\begin{equation}
\left\{
\begin{array}{rcl}
\delta C^{(1)} & = & -m\Lambda^{(1)}\, ,\\
& &  \\
\delta B & = & d\Lambda^{(1)}\, ,\\
\end{array}
\right.
\hspace{2cm}
\left\{
\begin{array}{rcl}
\delta B^{(6)} & = & m\Lambda^{(6)}\, ,\\
& & \\
\delta C^{(7)} & = & d\Lambda^{(6)}\, ,\\ 
\end{array}
\right.
\end{equation}

\noindent using which one can completely eliminate $C^{(1)}$
and $B^{(6)}$ everywhere. Then, the kinetic terms of these
St\"uckelberg fields become mass terms for $B$ and $C^{(7)}$
respectively. Only these two terms are massive.

Now, solutions describing $p$-branes in massive type~IIA SUEGRA
are automatically solutions describing the intersection of those
$p$-branes with a D8-brane associated to the mass parameter. General
solutions for the intersection of $p_{1}$ and $p_{2}$ branes have been
found in the literature using a generic model whose action contains
only kinetic terms for the dilaton and the $(p_{1}+2)$- and
$(p_{2}+2)$-form field strengths. Therefore, those solutions can
potentially describe correctly intersections involving a D8-brane and
a D0-, D2- and D4-brane, associated to massless fields. However, they
cannot correctly describe the intersections of a D8-brane and a
fundamental string, a solitonic 5-brane or a D6-brane. Study of the
supersymmetry algebra reveals that these solutions should exist and
preserve $1/4$ or the supersymmetries \cite{kn:Sa}. In the next two
sections we will present the corresponding solutions and will comment
on some of their unusual features.

%%%%%%%%%%%%%%%%%%%%%%%%%%%%%%%%%%%%%%%%%%%%%%%%%%%%%%%%%%%%%%%%%%%%%%

\section{Massive String}

This solution is given by

\begin{equation}
\left\{
\begin{array}{rcl}
ds^{2} & = & \Omega^{-1} \left( dt^{2} -dy^{2}\right) 
-d\vec{x}_{8}^{\, 2}\, ,\\
& & \\
B_{\underline{ty}} & = & \pm\left( \Omega^{-1}-1\right)\, ,\\
& & \\
C^{(1)}{}_{\underline{t}} & = & \pm my\, ,\\
& & \\
e^{-2\phi} & = & \Omega\, ,\\
\end{array}
\right.
\end{equation}

\noindent where 

\begin{equation}
\left\{
\begin{array}{rcl}
\vec{x}_{8} & = & \left(x^{1}, \ldots,x^{8} \right)
=\left(x^{m} \right)\, ,\\
& & \\
\partial_{\underline{m}}\partial_{\underline{m}}\Omega & = & -m^{2}\, ,\\
& & \\
\partial_{\underline{y}} \Omega = \alpha m\, .\\
\end{array}
\right.
\end{equation}

This solution has the following properties: 

\begin{enumerate}

\item The function $\Omega$ consists of three pieces: a piece linear in
$y$ (which is interpreted as the coordinate along the string and
perpendicular to the D8-brane), a piece quadratic in $\vec{x}_{8}$
(which are interpreted as the worldvolume coordinates of the D8-brane,
orthogonal to the string) and a harmonic function of $\vec{x}_{8}$:

\begin{equation}
\Omega= \alpha m y -\sum_{p}M_{p}x^{p}x^{p} +H(\vec{x}_{8})\, ,
\hspace{.5cm}
\sum_{p}M_{p}={\textstyle\frac{1}{2}}m^{2}\, ,
\hspace{.5cm}
\partial_{\underline{m}}\partial_{\underline{m}} H=0\, .
\end{equation}

Thus, it can describe, in principle, several objects in equilibrium.

\item In the massless limit $\Omega(y,\vec{x}_{8})=H(\vec{x}_{8})$ and
  for the right choice of $H$ it is just the fundamental string
  solution \cite{kn:DGHR}.

\item The limit in which the string is eliminated is unattainable from
this solution. Even if we set $H=0$ $\Omega$ is still non-trivial and
the solution will have only $1/4$ of the supersymmetries unbroken. 
  
\item The $C^{(1)}$ field can be completely gauged away, canceling the
  $\mp 1$ in $B_{\underline{ty}}$. We have introduced it in order to
  have $B_{\underline{ty}}$ in the form which corresponds to a
  fundamental string source. (It can be argued that there is a
  D0-brane in the intersection between the string and the D8-brane, as
  we will see when we study the unbroken supersymmetry the solution.)
  
\item We have a solution for any value of the constant $\alpha$.
  However, only for $\alpha=\mp 1$ the solution is supersymmetric.
  This is a quite unusual behavior

\end{enumerate}

Let us now find the unbroken supersymmetries. We will only analyze the
dilatino supersymmetry rule to show how it works. In this case

\begin{equation}
\delta_{\epsilon}\lambda =   
\left(\not\!\partial\phi
+{\textstyle\frac{1}{2\cdot 3!}}\Gamma_{11}\not\!\! H\right)\epsilon 
+{\textstyle\frac{5i}{4}} m e^{\phi} \epsilon
-{\textstyle\frac{3i}{8}} e^{\phi} \not\! G^{(2)} \Gamma_{11}\epsilon\, ,
\end{equation}

\noindent with 

\begin{equation}
\left\{
\begin{array}{rcl}
\not\!\! H & = & \mp 3! 
\partial_{\underline{m}} \Omega \Gamma^{m} \Gamma^{0y}\, ,\\
& & \\
\not\!\partial\phi & = & -{\textstyle\frac{1}{2}} \Omega^{-1} 
\partial_{\underline{m}} \Omega \Gamma^{m}
-{\textstyle\frac{1}{2}} \Omega^{-1/2} 
\partial_{\underline{y}} \Omega \Gamma^{y}\, ,\\
& & \\
\not\! G^{(2)} & = & \pm 2m \Gamma^{0y}\, .\\
\end{array}
\right.
\end{equation}

Substituting into the dilatino supersymmetry rule we find

\begin{equation}
-{\textstyle\frac{1}{2}} \Omega^{-1} 
\partial_{\underline{m}} \Omega \Gamma^{m}
\left[1\mp\Gamma^{0y}\Gamma_{11}\right]  \epsilon
-{\textstyle\frac{1}{2}} \Omega^{-1/2} 
\partial_{\underline{y}} \Omega \Gamma^{y} \epsilon
+{\textstyle\frac{i}{4}}m \Omega^{-1/2} 
\left[ 5 \mp 3\Gamma^{0y}\Gamma_{11}\right] \epsilon=0\, .
\end{equation}

\noindent The first term cancels if we impose 

\begin{equation}
{\textstyle\frac{1}{2}}
\left[1\mp\Gamma^{0y}\Gamma_{11}\right]  \epsilon =0\, ,
\end{equation}

\noindent which is the condition satisfied by the Killing spinor of the
fundamental string. This operator is a projector and therefore has
eigenvalues $1$ or $0$. The trace is $16$, one half of the trace of the
identity and therefore this condition breaks a half of the
supersymmetries.  Using this condition also in the third term we get

\begin{equation}
-\partial_{\underline{y}} \Omega \Gamma^{y} \epsilon
+im \epsilon=0\, ,  
\end{equation}

\noindent which is solved by $\alpha=\mp 1$ and

\begin{equation}
m{\textstyle\frac{1}{2}}
\left[1\mp i\Gamma^{y}\right]  \epsilon =0\, ,
\end{equation}

\noindent which is the condition satisfied by the Killing
spinor of a D8-brane. For analogous reasons, this second condition
breaks a half of the supersymmetries for $m\neq 0$. These two
projectors commute and therefore both conditions can be fulfilled
simultaneously. Since the trace of the product of both projectors
is $8$, $1/4$ of the supersymmetries are preserved.

The gravitino equation also vanishes if the Killing spinor is 

\begin{equation}
\epsilon =\Omega^{1/4}\epsilon_{0}\, ,
\end{equation}

\noindent where $\epsilon_{0}$ is a constant spinor satisfying the
above constraints.

Now, observe that if the Killing spinor is an eigenspinor of the 
fundamental string and D8-brane projectors, then it obeys automatically

\begin{equation}
{\textstyle\frac{1}{2}}
\left[1\mp i\Gamma^{0}\Gamma_{11}\right]  \epsilon =0\, ,
\end{equation}

\noindent which is the condition of the D0-brane Killing spinor.  This
may seem a bit surprising since $C^{(1)}$ is trivial (unless $y$ is a
compact coordinate).  However, its field strength $G^{(2)}$, which is
the meaningful quantity is not trivial. 

For all these reasons one can identify this solution with the
intersection of fundamental string and a D8-brane over a
D0-brane\footnote{A string solution to a class of massive supergravity
  theories was recently given in \cite{kn:Si}. However, the mass
  parameter in that model is of NSNS type and there is no mass term
  for $B$. Thus it cannot describe the fundamental string of the
  massive type~IIA theory.}.

%%%%%%%%%%%%%%%%%%%%%%%%%%%%%%%%%%%%%%%%%%%%%%%%%%%%%%%%%%%%%%%%%%%%%%

\section{Massive D6-Brane}

This solution is given by

\begin{equation}
\left\{
\begin{array}{rcl}
ds^{2} & = & \Omega^{-1/2} \left( dt^{2} -d\vec{y}_{6}^{\, 2}\right) 
-\Omega^{1/2}d\vec{x}_{3}^{\, 2}\, ,\\
& & \\
B_{\underline{mn}} & = & \mp \frac{m}{3}\epsilon_{mnp}x^{p}\, ,\\
& & \\
B^{(6)}{}_{\underline{t}\underline{y^{2}}\cdots\underline{y^{6}}}
& = & \pm m y^{1}\, ,\\
& & \\
C^{(7)}{}_{\underline{t}\underline{y^{1}}\cdots \underline{y^{6}}} 
& = & \pm \left(\Omega^{-1}-1 \right)\, ,\\
& & \\
e^{-2\phi} & = & \Omega^{3/2}\, ,\\
\end{array}
\right.
\end{equation}

\noindent where 

\begin{equation}
\left\{
\begin{array}{rcl}
\vec{y}_{6} & = & \left(y^{1},\ldots,y^{6} \right) 
=\left( y^{i}\right)\, ,\\
& & \\
\vec{x}_{3} & = & \left(x^{1},x^{2},x^{3}\right)
=\left(x^{m} \right)\, ,\\
& & \\
\partial_{\underline{m}}\partial_{\underline{m}}\Omega & = & -m^{2}\, ,\\
& & \\
\partial_{\underline{y^{1}}} \Omega & = & \alpha m\, .\\
\end{array}
\right.
\end{equation}

Some remarks are necessary:
\begin{enumerate}
  
\item As $C^{(1)}$ in the massive string case, $B^{(6)}$ is pure
  gauge but we have introduced it only for the sake of consistency.

\item $B$ is {\it not pure} gauge. A non-trivial $C^{(7)}$ (necessary
for a D6-brane) implies a non-trivial $H^{(7)}$ and, by Hodge duality, a
nontrivial $H$ and a non-trivial $B$. This (plus the constraints
of unbroken supersymmetry) will give support to the interpretation
that there is a solitonic 5-brane in the intersection.

\item The coordinate $y^{1}$ has been chosen for simplicity but any
other direction in the D6-brane worldvolume (coordinates
$\left(t,y^{i}\right)$) would do as direction orthogonal to the
solitonic 5-brane and D8-brane. 

\item Again, the function $\Omega$ consists of three pieces: a piece
linear in $y^{1}$ (the coordinate orthogonal to the solitonic 5-brane
and the D8-brane), a piece quadratic in $\vec{x}_{3}$ (which are
interpreted as worldvolume coordinates of the D8-brane, orthogonal to
both the solitonic 5-brane and the D8-brane) and a harmonic function of
$\vec{x}_{3}$:

\begin{equation}
\Omega= \alpha m y^{1} -\sum_{p}M_{p}x^{p}x^{p}
 +H(\vec{x}_{3})\, ,
\hspace{.5cm}
\sum_{p}M_{p}={\textstyle\frac{1}{2}}m^{2}\, ,
\hspace{.5cm}
\partial_{\underline{m}}\partial_{\underline{m}} H=0\, .
\end{equation}

Thus, it can describe, in principle, several objects in equilibrium.

\item In the massless limit $\Omega(y^{1},\vec{x}_{3})=H(\vec{x}_{3})$
  and the right choice of $H$ it is just the D6-brane solution.
  
\item We have a solution for any value of the constant $\alpha$.
However, only for $\alpha=\mp 1$ the solution is supersymmetric.
Actually, one finds that (for those values of $\alpha$) the Killing
spinor is 

\begin{equation}
\epsilon=\Omega^{-1/8} \epsilon_{0}\, ,  
\end{equation}

\noindent where $\epsilon_{0}$ is a constant spinor which satisfies

\begin{equation}
\left\{
\begin{array}{rcl}
\frac{1}{2}\left(1\mp i \Gamma^{01\cdots 6} \right) \epsilon_{0} & = & 0\, ,\\
& & \\
m{\textstyle\frac{1}{2}}
\left[1\mp i\Gamma^{y}\right]  \epsilon_{0} & = & 0\, .\\
\end{array}
\right.
\end{equation}

If both equations are satisfied, then the following equation is
satisfied

\begin{equation}
{\textstyle\frac{1}{2}}
\left[1\pm \Gamma^{02\cdots 6}\right]  \epsilon =0\, ,
\end{equation}

\noindent which is the condition satisfied by the solitonic 5-brane 
Killing spinor.

\end{enumerate}

It is reasonable to identify these solution with the intersection of
a D6- and D8-brane over a solitonic 5-brane.

%%%%%%%%%%%%%%%%%%%%%%%%%%%%%%%%%%%%%%%%%%%%%%%%%%%%%%%%%%%%%%%%%%%%%%

\section{Conclusion}

We have presented two new supersymmetric solutions of massive type~IIA
SUEGRA which have interpreted as the intersection of a string and
a D6-brane with a D8-brane over a D0-brane and a solitonic 5-brane
respectively. They have a number of remarkable features some of them
quite unusual: the massless limit exists (this is the limit in which
the D8-brane is removed) but the other limits do not exist. The fact
that they consist of 3-objects but $1/4$ of the supersymmetries is
preserved has also been observed in other contexts. Furthermore, the
string solutions is completely localized and depends on all coordinates
except on time. 

It would be interesting to find other solutions of massive type~IIA
SUEGRA and to study their possible 11-dimensional origin.  Work on
this direction is in progress.

%%%%%%%%%%%%%%%%%%%%%%%%%%%%%%%%%%%%%%%%%%%%%%%%%%%%%%%%%%%%%%%%%%%%%%%%%

\section*{Acknowledgments}

The authors wish to thank E.~\'Alvarez, H.J.~Boonstra, R.~Empar\'an
for fruitful discussions. The work of P.M.~has been partially
supported by the European Union under contract number
ERBFMBI-CT96-0616.  The work of B.J.~and T.O.~has been supported by
the European Union TMR program FMRX-CT96-0012 {\sl Integrability,
  Non-perturbative Effects, and Symmetry in Quantum Field Theory}. The
work of T.O.~has also been supported by the Spanish grant AEN96-1655.

%%%%%%%%%%%%%%%%%%%%%%%%%%%%%%%%%%%%%%%%%%%%%%%%%%%%%%%%%%%%%%%%%%%%%%%

\end{document}